\documentstyle[12pt]{article}
\begin{document}
\large

\begin{center}{\large\bf ON THE PROBLEM OF INTERACTIONS IN QUANTUM
THEORY}\end{center}
\vskip 1em \begin{center} {\large Felix M. Lev} \end{center}
\vskip 1em \begin{center} {\it Laboratory of Nuclear
Problems, Joint Institute for Nuclear Research, Dubna, Moscow region
141980 Russia (E-mail:  lev@nusun.jinr.ru)} \end{center}
\vskip 1em

\begin{abstract}
The structure of representations describing systems of free
particles in the theory with the invariance group SO(1,4) is
investigated. The property of the particles to be free means as usual
that the representation describing a many-particle system is the
tensor product of the corresponding single-particle representations
(i.e. no interaction is introduced). It is shown that the mass
operator contains only continuous spectrum in the interval
$(-\infty,\infty)$ and such representations are unitarily equivalent to
ones describing interactions (gravitational, electromagnetic
etc.). This means that there are no bound states in the theory and
the Hilbert space of the many-particle system contains a subspace of
states with the following property: the action of free representation
operators on these states is manifested in the form of different
interactions. Possible consequences of the results are discussed.
\end{abstract}

\vskip 0.5em

\section{General remarks on quantum theories}
\label{S1}

  The existing quantum theories are usually based on the following
procedure: the Lagrangian of the system under consideration is
written as $L=L_m+L_g+L_{int}$ where $L_m$ is the Lagrangian of
"matter", $L_g$ is the Lagrangian of gauge fields and $L_{int}$ is
the interaction Lagrangian. The symmetry conditions do not define
$L_{int}$ uniquely since at least the interaction constant is
arbitrary. Nevertheless such an approach has turned out to be
highly successful in QED, electroweak theory and QCD. At the same
time the difficulties in constructing quantum gravity have not
been overcome.

 The popular idea expressed by many physicists is that such notions
as matter and interactions are not fundamental --- they are only
manifestations of some properties of space-time (see e.g. Ref.
\cite{Wheeler}). On the other hand the problem arises what is the
meaning of space-time on the quantum level.

Indeed, there is no operator corresponding to time (see e.g.
the discussion in Ref. \cite{time})
and the latter is considered only as a classical
quantity. Moreover, it has become clear already in 30th that in
relativistic quantum theory there is no operator having all the
properties of the coordinate operator (see e.g. Ref. \cite{Heg}).
As a consequence, the
quantity $x$ in the Lagrangian density $L(x)$ is not the coordinate
in Minkowski space but some parameter which becomes the coordinate
only in the classical limit.

 One can consider the Lagrangian only as an auxiliary tool for
constructing all the generators of the symmetry group in the
framework of the canonical formalism (see e.g. Ref. \cite{Wein}).
Nevertheless the practical realization of such an approach
encounters serious mathematical problems. One of the reason is
that the interacting field operators can be treated only
as operator valued distributions \cite{BLOT} and therefore the
product of two local field operators at coinciding points is not
well defined. The problem of the correct definition of such products
is known as the problem of constructing composite operators (see e.g. Ref.
\cite{Zim}). So far this problem has been solved only in the framework
of perturbation theory for special models. When perturbation theory
does not apply the usual prescriptions are to separate the arguments
of the operators in question and to define the composite operator as
a limit of nonlocal operators when the separation goes to zero (see e.g.
Ref. \cite{J} and references therein).
 However the Lagrangian contains products of local operators at the
same point. If one separates the arguments of these
operators, the Lagrangian immediately becomes nonlocal and it is not
clear how to use the Noether procedure in this case.

As a consequence, the generators constructed in a standard way are
usually not well-defined
(see e.g. Ref. \cite{JL}) and the theory contains anomalies.
It is also worth noting that there exists the well-known paradox:
on the one hand the renormalizable perturbation theory, which is
formally based on interaction picture, is in beautiful agreement
with many experimental data, while on the other hand,
according to the Haag theorem \cite{Haag,BLOT}, the interaction
picture does not exist.

 One of the advantages of superstring theories is that they
involve products of operators at different points and there
exist theories without anomalies \cite{GSW}. At the same time
the theories involve such notions as Lagrangian, interaction and
perturbation theory.

 In chiral theories (see e.g. Ref. \cite{Per} and references
therein) the interaction Lagrangian is usually not introduced
but the fields have the range in some nonlinear manifold. There
exist serious problems in quantizing such theories.

 The above remarks show that at present physicists have no clear
answer to the questions whether in quantum theory the notions of
space-time, Lagrangian and interaction are primary or they are
manifestations of some deeper reasons. According to the Heisenberg
program (which was very popular in 60th) such notions are rudiments
which will not survive in the future theory (see e.g. the
discussion in Ref. \cite{BLP}). According to this program, the only
fundamental operator is the $S$-operator defined on the
Hilbert space of elementary particles while bound states are
manifested only as poles in the matrix elements of this operator.
It is also interesting
to pose the following problem. Is it possible that the ultimate physical
theory is fully defined by the choice of the symmetry group and
there is no need to introduce the Lagrangian and interactions?

 In the present paper we consider a model which in our opinion
can shed light on this problem. We choose the de Sitter group
SO(1,4) (more precisely its covering group
$\overline{SO(1,4)}$) as the symmetry group. We require as usual
that the elementary particles are described by unitary irreducible
representations (UIRs) of this group and different realizations of
such representations are described in Sect. \ref{S2}. Then we
assume that the representation describing the many-particle system
is the tensor product of the corresponding single-particle
representations. According to the usual philosophy this means
that the particles are free and no interaction is introduced. In
Sect. \ref{S3} we explicitly calculate the free many-particle mass
operator and show that the spectrum of this operator contains the
whole interval $(-\infty,\infty)$. As shown in Sect. \ref{S4}, such
an operator is unitarily equivalent to the mass operator containing
interactions (gravitational, strong, electromagnetic etc.). Finally,
Sect. \ref{S5} is discussion.

  It is worth noting that the unusual properties of SO(1,4)-invariant
theories considered in this paper are specific only for these
theories while SO(2,3)-invariant theories have many common features
with Poincare-invariant ones (the mass of the elementary particle
coincides with the minimal value of its energy, the mass of
the two-particle system has the minimal value $m_1+m_2$ etc.
\cite{Fron}).

\begin{sloppypar}
\section{Realizations of single-particle representations of
the SO(1,4) group}
\label{S2}
\end{sloppypar}

 The de Sitter group SO(1,4) is the symmetry group of the
four-dimensional manifold  which can be described as follows.
If $(x_0,x_1,x_2,x_3,x_4)$ are the coordinates in the
five-dimensional space, the manifold is the set of points
satisfying the condition
\begin{equation}
x_0^2 -x_1^2 -x_2^2 -x_3^2-x_4^2 =-R_0^2
\label{1}
\end{equation}
where $R_0>0$ is some parameter. Elements of a map of the point
$(0,0,0,0,R_0)$ (or $(0,0,0,0,-R_0)$) can be parametrized by the
coordinates $(x_0,x_1,x_2,x_3)$. If $R_0$ is very large then such a
map  proceeds
to Minkowski space and the action of the de Sitter group on this map
--- to  the action of the Poincare group.
The quantity $R_0^2$ is often written as $R_0^2=3/\Lambda$ where
$\Lambda$ is the cosmological constant. The existing astronomical
data show that $\Lambda$ is very small and the usual estimates based
on popular cosmological models give $R_0>10^{26}cm$. On the other
hand, in models based on the de Sitter cosmology the quantity
$R_0$ is related to the Hubble constant $H$ as $R_0=1/H$ and in
this case $R_0$ is of order $10^{27}cm$ \cite{desitt}.

\begin{sloppypar}
The representation generators of the SO(1,4) group
$L^{ab}$ ($a,b=0,1,2,3,4$, $L^{ab}=-L^{ba}$) should satisfy the
commutation relations
\begin{equation}
[L^{ab},L^{cd}]=-\imath (\eta^{ac}L^{bd}+\eta^{bd}L^{as}-
\eta^{ad}L^{bc}-\eta^{bc}L^{ad})
\label{2}
\end{equation}
where $\eta^{ab}$ is the diagonal metric tensor such that
$\eta^{00}=-\eta^{11}=-\eta^{22}=-\eta^{33}=-\eta^{44}=1$.
\end{sloppypar}

 In conventional quantum theory elementary particles are described
by UIRs of the symmetry group or its Lie algebra.
 If one assumes that the role of the symmetry group is played by
the Poincare group, then the representations are described by
ten generators --- six generators of the Lorentz group and the
four-momentum operator. In the units $c=\hbar =1$ the former are
dimensionless while the latter has the dimension $(length)^{-1}$.
If however the symmetry group is the de Sitter group SO(1,4), then
all the generators in the units $c=\hbar =1$
are dimensionless. There exists wide literature devoted to the
UIRs of this group
(see e.g. Refs. \cite{Str,Men,Moy,lev}). Below we describe three
different realizations of the UIRs. The reader can explicitly
verify that the generators indeed satisfy Eq. (\ref{2}).

\begin{sloppypar}
If $s$ is the spin of the particle under consideration, then we
use $||...||$ to denote the norm in the space of the
UIR of the group SU(2) with the spin $s$. Let $v=(v_0=(1+{\bf v}^2)^{1/2},
{\bf v})$ be the element of the Lorentz hyperboloid of
four-velocities and $dv$ be the Lorentz invariant volume element
on this hyperboloid. Then one can realize the UIR under
consideration in the space of functions $\{f_1(v),f_2(v)\}$ on
two Lorentz hyperboloids with the range in the space of the
UIR of the group SU(2) with the spin $s$ and such that
\begin{equation}
\int\nolimits [||f_1(v)||^2+||f_2(v)||^2]dv <\infty
\label{3}
\end{equation}
The  explicit  calculation  shows  that  the  action  of the
generators on $f_1(v)$ has the form
\begin{eqnarray}
&&{\bf M}=l({\bf v})+{\bf s},\quad {\bf N}==-\imath v_0
\frac{\partial}{\partial {\bf v}}+\frac{{\bf s}\times {\bf v}}
{v_0+1}, \nonumber\\
&& {\bf B}=\mu {\bf v}+\imath [\frac{\partial}{\partial {\bf v}}+
{\bf v}({\bf v}\frac{\partial}{\partial {\bf v}})+\frac{3}{2}{\bf v}]+
\frac{{\bf s}\times {\bf v}}{v_0+1},\nonumber\\
&& L_{04}=\mu v_0+\imath v_0({\bf v}
\frac{\partial}{\partial {\bf v}}+\frac{3}{2})
\label{4}
\end{eqnarray}
where ${\bf M}=\{L^{23},L^{31},L^{12}\}$,
${\bf N}=\{L^{01},L^{02},L^{03}\}$,
${\bf B}=-\{L^{14},L^{24},L^{34}\}$, ${\bf s}$ is the spin operator,
and ${\bf l}({\bf v})=-\imath{\bf v}
\times \partial/\partial {\bf v}$.
 The action of  $L^{ab}$    on  $f_2(v)$  is
obtained from Eq. (\ref{4}) by the substitution $\mu \rightarrow -\mu$.
\end{sloppypar}

 Such a realization is used to obtain the
possible closest analogy between the  representations  of the SO(1,4)
and the Poincare group. It is easy to see that the operators ${\bf M}$
and ${\bf N}$ in Eq. (\ref{4}) have the
same form as  for  the  standard  realization  of the single-particle
representations of the Poincare group (see e.g. Ref. \cite{Nov,lev1})
while the contraction of the representation  Eq. (\ref{4}) into  the
standard realization of the Poincare  group  is  accomplished  as
follows. Denote $m=\mu/R$, ${\bf P}={\bf B}/R$ and $E=L_{04}/R$
where $R>0$ is some quantity of order $R_0$ (but not necessarily
equal to $R_0$). Consider the action of the generators on functions
which do not depend on $R$ in the usual system of units. Then, as
follows from  Eqs. (\ref{1}) and (\ref{4}), in the limit
$R\rightarrow \infty$ we obtain the standard representation of the
Poincare group for a particle with the mass $m$, since ${\bf P}=
m{\bf v}$,  $E=mv_0$
(in this case one has the representation  of  the  Poincare  group
with the negative energy on the second hyperboloid).

 Since the representation generators of the SO(1,4) group are
dimensionless (in the units $c=\hbar =1$), any quantal description
in the SO(1,4)-invariant theory involves only dimensional
quantities. In particular, as seen from Eq. (\ref{4}), the quantal
description of particles in such a theory does not involve any
information about the quantity $R_0$ (this property is clear from
the fact that the elements of the SO(1,4) group describe only
homogeneous transformations of the manifold defined by Eq. (\ref{1})).
The latter comes into play only when we wish to interpret the
results in terms of quantities used in the Poincare-invariant theory.
Therefore if we assume that de Sitter invariance is fundamental
and Poincare invariance is only approximate, it is reasonable
to think that the de Sitter masses $\mu$ are fundamental while
the quantities $m$, $R_0$ and $R$ are not (see also the discussion
in Ref. \cite{LM}). In particular, $R$ (or even $R_0$) may be
time-dependent (see below) and in this case the usual masses will
be time-dependent too.

It is also possible to realize the UIR in the space of functions
$\varphi (u)$
on the  three-dimensional unit sphere $S^3$ in the four-dimensional
space with the range in the space of the UIR of the group SU(2) with
the spin $s$ and such that
\begin{equation}
\int\nolimits ||\varphi(u)||^2du <\infty
\label{5}
\end{equation}
where $du$ is the SO(4) invariant volume element on $S^3$.
Elements  of $S^3$ can
be represented as $u=({\bf u},u_4)$ where $u_4=\pm (1-{\bf u}^2)^{1/2}$
for the upper and lower hemispheres respectively. Then the
explicit calculation  shows  that  the  generators for  this
realization have the form
\begin{eqnarray}
&&{\bf M}=l({\bf u})+{\bf s},\quad {\bf B}=\imath u_4
\frac{\partial}{\partial {\bf u}}-{\bf s}, \nonumber\\
&& {\bf N}=\imath [\frac{\partial}{\partial {\bf u}}-
{\bf u}({\bf u}\frac{\partial}{\partial {\bf u}})]
-(\mu +\frac{3\imath}{2}){\bf u}+{\bf u}\times {\bf s}-u_4{\bf s},\nonumber\\
&& L_{04}=(\mu +\frac{3\imath}{2})u_4+\imath u_4{\bf u}
\frac{\partial}{\partial {\bf u}}
\label{6}
\end{eqnarray}

  Since Eqs. (\ref{3}) and (\ref{4}) on the one hand and
Eqs. (\ref{5}) and (\ref{6}) on  the
other  are  the  different  realization  of  one  and   the   same
representation, then there exists a unitary operator transforming
functions $f(v)$ into $\varphi (u)$ and operators (\ref{4}) into
operators (\ref{6}). For example in the spinless case
\begin{equation}
\varphi (u)=exp(-\frac{\imath}{2}\,\mu \,lnv_0)v_0^{3/2}f(v)
\label{7}
\end{equation}
where ${\bf u}=-{\bf v}/v_0$. In view of this relation, the
sphere $S^3$ is usually interpreted in the literature as the
velocity space (see e.g. Refs. \cite{Men,Moy}), but, as argued
in Ref. \cite{lev}, there are serious arguments to interpret
$S^3$ as the coordinate space. Below we give additional arguments
in favor of this point of view.

 As follows from Eq. (\ref{1}), if $x_0$ is fixed then the set
of the points satisfying this relation is the three-dimensional
sphere $S^3(R_1)$ with the radius $R_1=(R_0^2+x_0^2)^{1/2}$.
This sphere is invariant under the action of the SO(4) subgroup
of the SO(1,4) group. The operators ${\bf B}$ and ${\bf M}$ are
the representation generators of the SO(4) subgroup. We can choose
${\bf x}=R_1{\bf u}$ as the coordinates on $S^3(R_1)$. In these
coordinates the operators ${\bf B}$ and ${\bf M}$ given by Eq.
(\ref{6}) are the generators of the representation of the group
of motions of $S^3(R_1)$ induced from the representation of the
SO(3) subgroup with the generators ${\bf s}$.

 Consider a vicinity of the south pole of $S^3(R_1)$ such that
$|{\bf x}|\ll R_1$. Then the generators in Eq. (\ref{6}) have the form
\begin{eqnarray}
{\bf B}=R_1{\bf p},\quad {\bf M}={\bf l}({\bf x})+{\bf s},\quad
{\bf N}=-R_1{\bf p}, \quad L_{04}=mR_1
\label{8}
\end{eqnarray}
where ${\bf p}=-\imath \partial/\partial {\bf x}$ and $m=\mu/R_1$.
Therefore ${\bf B}/R_1$ is the de Sitter analog of the momentum
operator, but $L_{04}/R_1$ in this realization is not the de Sitter
analog of the energy operator.

 The reason of such a situation is as follows. In Poincare
invariant theories one can consider wave functions defined on the
conventional three-dimensional space $R^3$. The operators ${\bf M}$
and ${\bf P}$ are the representation generators of the group of
motions of $R^3$. From the remaining generators, $E$ and ${\bf N}$,
only $E$ commutes with ${\bf M}$ and ${\bf P}$. For this reason
$E$ can be chosen as the operator responsible for the evolution
of the system under consideration while stationary states are
the eigenstates of $E$. In the SO(1,4) case one can
consider wave functions defined on $S^3(R_1)$ at different values
of $x_0$. However none of the generators $L_{04},{\bf N}$ commutes
with all the operators ${\bf M}$ and ${\bf B}$. At the same time
the operator $E_{dS}=(L_{04}^2+{\bf N}^2)^{1/2}$ satisfies this
property. Hence $E_{dS}$ can be treated as the operator responsible
for the evolution. At the conditions described by Eq. (\ref{8}),
$E_{dS}=R_1(m^2+{\bf p}^2)^{1/2}$ and therefore $E_{dS}$ can be
considered as the de Sitter analog of the energy operator.

 The inconvenience of working with ${\bf B}$ as the de Sitter
analog of the momentum operator is that different components of
${\bf B}$ commute with each other only when $R_1\rightarrow
\infty$. We can define the operators
${\bf Q}_+=(L_{1+},L_{2+},L_{3+})$ and ${\bf Q}_-=(L_{1-},L_{2-},L_{3-})$,
where the $\pm$ components of five-vectors are defined as
$x_{\pm}=x_4\pm x_0$. Then, as follows from Eq. (\ref{2}),
different components of ${\bf Q}_+$ commute with each other and
the same is valid for ${\bf Q}_-$.

 It is easy to see that $2{\bf u}/(1-u_4)$ is the stereographic
projection of the point $({\bf u},u_4)\in S^3$ onto the
three-dimensional space. Now we will use ${\bf x}$ to denote the
quantity ${\bf x}=2R{\bf u}/(1-u_4)$ where $R$ is some quantity
of order $R_0$ and $R=R_1$ is a reasonable choice. In the space of
functions $\varphi ({\bf x})$ on $R^3$ with the range in the space
of the UIR of the group SU(2) with the spin $s$ and such that
\begin{equation}
\int\nolimits ||\varphi({\bf x})||^2d^3{\bf x} <\infty
\label{9}
\end{equation}
the generators of the UIR of the SO(1,4) group can be realized as
follows
\begin{eqnarray}
&&{\bf M}={\bf l}({\bf x})+{\bf s}, \quad L_{+-}=-2(\mu+{\bf x}{\bf p})+
3\imath,\quad {\bf Q}_+=-2R{\bf p},\nonumber\\
&& {\bf Q}_-=\frac{1}{2R}[-2\mu {\bf x}+{\bf x}^2{\bf p}-
2{\bf x}({\bf x}{\bf p})+3\imath {\bf x} +2({\bf s}\times {\bf x})]
\label{10}
\end{eqnarray}
where we again use ${\bf p}=-\imath \partial/\partial {\bf x}$, but
the quantities ${\bf x}$ and ${\bf p}$ are not the same as in
Eq. (\ref{8}).

 The UIR realized by Eqs. (\ref{4}), (\ref{6}) or (\ref{10})
belongs to the so called principal series. It can be characterized
by the condition that $\mu^2\geq 0$, i.e. $\mu$ is real. In
contrast with the UIRs of the Poincare group, the sign of $\mu$
does not make it possible to distinguish the UIRs describing
particles and antiparticles (see e.g. the discussion in Ref.
\cite{Men}), and the UIRs with $\mu$ and $-\mu$ are unitarily
equivalent.

 The Casimir operator of the second order can be written as
\begin{eqnarray}
&&I_2=-\sum_{ab} L_{ab}L^{ab}=\frac{1}{2}(L_{+-})^2-\nonumber\\
&&({\bf Q}_+{\bf Q}_-+{\bf Q}_-{\bf Q}_+)-2{\bf M}^2
\label{11}
\end{eqnarray}
A direct calculation shows that in the case described by
Eqs. (\ref{4}), (\ref{6}) and (\ref{10})
\begin{equation}
I_2=2(\mu^2-{\bf s}^2+\frac{9}{4})
\label{12}
\end{equation}

 In Poincare invariant theories the spectrum of the mass operator
can be defined as the spectrum of the energy operator in the
subspace of states with zero total momentum. As follows from
Eq. (\ref{10}), for the UIRs of the SO(1,4) group corresponding
to the principal series, the spectrum of the mass operator can
be defined from the condition that the action of $L_{+-}$ on the
states with zero momentum is equal to $-2\mu+3\imath$. The
presence of $3\imath$ in this expression does not contradict
the Hermiticity of $L_{+-}$ since $L_{+-}$ does not commute with
${\bf Q}_+$.

\begin{sloppypar}
\section{Free many-particle mass operator in the SO(1,4)-invariant
theory}
\label{S3}
\end{sloppypar}

  The representation describing a system of $N$ noninteracting
particles is constructed as the tensor product of  the  corresponding
single-particle representations, and the representation generators are
equal to the sums of single-particle  generators,  i.e.
\begin{equation}
L_{ab}=\sum_{n=1}^{N}L_{ab}^{(n)}
\label{13}
\end{equation}
where $L_{ab}^{(n)}$ are the generators for the $n$th particle. Each
generator acts through the variables of its "own"
particle, as described in the preceding section, and through the
variables of other particles it acts as the identity operator.

The tensor product of single-particle representations can be
decomposed into the direct integral of UIRs  and there exists the well
elaborated general theory \cite{Dix}. In  the  given case, among the
UIRs there may be not only representations of the principal
series but also UIRs of other series.

 We first consider the case of two particles 1 and 2. Suppose that the
UIRs for them are realized as in Eq. (\ref{4}). Introduce the
conventional masses and momenta $m_j=\mu_j/R$, ${\bf p}_j=m_j{\bf v}_j$
($j=1,2$). We can define the variables describing the system as a whole
and the internal variables. The usual nonrelativistic variables are:
\begin{equation}
{\bf P}={\bf p}_1+{\bf p}_2,\quad {\bf k}=\frac{m_2{\bf p}_1-
m_1{\bf p}_2}{m_1+m_2}
\label{14}
\end{equation}
Then, in the approximation when the particle velocities are very small,
it follows from Eqs. (\ref{4}) and (\ref{13}) that the two-particle
generators have the form
\begin{eqnarray}
&& {\bf M}={\bf l}({\bf P})+{\bf S},\quad {\bf N}=-\imath (m_1+m_2)
\frac{\partial}{\partial {\bf P}}, \nonumber\\
&&L_{04}=R(m_1+m_2)+\imath({\bf k}\frac{\partial}{\partial {\bf k}}+
\frac{3}{2})+\imath ({\bf P}\frac{\partial}{\partial {\bf P}} +
\frac{3}{2}),\nonumber\\
&& {\bf B}=R{\bf P}+\imath (m_1+m_2) \frac{\partial}{\partial {\bf P}}
\label{15}
\end{eqnarray}
where ${\bf S}={\bf l}({\bf k})+{\bf s}_1+{\bf s}_2$. The comparison
of the expressions (\ref{4}) and (\ref{15}) for ${\bf M}$ shows that
${\bf S}$ plays the role of the spin operator for the system as a whole
(in full analogy with the conventional quantum mechanics).

  By analogy with Eq. (\ref{12}) we can define the mass operator
$M_{dS}$ for the system as a whole. Namely, if $I_2$ is the Casimir
operator for the system as a whole defined by this expression, then
\begin{equation}
I_2=2(M_{dS}^2-{\bf S}^2+\frac{9}{4})
\label{16}
\end{equation}
In turn, the conventional mass operator $M$ can be defined as
$M_{dS}/R$.

 As follows from Eqs. (\ref{15}) and (\ref{16}), for slow particles
in first order in $1/R$
\begin{equation}
M=m_1+m_2+\frac{\imath}{R}({\bf k}\frac{\partial}{\partial {\bf k}}+
\frac{3}{2})
\label{17}
\end{equation}
We shall see below that this expression is correct for any
velocities and in any order in $1/R$ if only the representations
of the principal series are taken into account.

 Equation (\ref{17}) means that for very slow particles the
de Sitter correction to the classical nonrelativistic Hamiltonian
is equal to $\Delta H_{nr}=({\bf k}{\bf r})/R$ where ${\bf r}=
{\bf x}_1-{\bf x}_2$ is the vector of the relative distance
between the particles (this quantity is conjugated with ${\bf k}$).
As follows from the classical equations of motion, $\ddot{{\bf r}}
={\bf r}/R^2$. Therefore the correction corresponds to the well-known
fact that in the classical SO(1,4)-invariant theory there exists the
antigravity, and the force of (cosmological) repulsion between
particles is
proportional to the distance between them. It is also interesting
to note that the de Sitter antigravity is in some sense even more
universal than the usual gravity: the force of repulsion does not
depend on the parameters characterizing the particles (even on
their masses).

 Now we again consider the case of two particles but suppose that the
UIRs for them are realized as in Eq. (\ref{10}). Introduce the
standard nonrelativistic variables
\begin{equation}
{\bf X}=\frac{m_1{\bf x}_1+m_2{\bf x}_2}{m_1+m_2}
=\frac{\mu_1{\bf x}_1+\mu_2{\bf x}_2}{\mu_1+\mu_2},\quad {\bf r}=
{\bf x}_1-{\bf x}_2
\label{18}
\end{equation}
Then a direct calculation of the two-particle generators gives:
\begin{eqnarray}
&& L_{+-}=-2(\mu_1+\mu_2+{\bf X}{\bf P})+
2\imath ({\bf r}\frac{\partial}{\partial {\bf r}}+3),\nonumber\\
&&{\bf Q}_+=-2R{\bf P}, \quad
{\bf M}={\bf l}({\bf X})+{\bf S},\nonumber\\
&& {\bf Q}_-=-(m_1+m_2){\bf X}+
\frac{1}{2R}[{\bf X}^2{\bf P}+
\frac{m_1m_2}{(m_1+m_2)^2}{\bf r}^2{\bf P}-\nonumber\\
&&2\imath ({\bf r}{\bf X}\frac{\partial}{\partial {\bf r}}-
\imath\frac{m_2-m_1}{m_1+m_2}{\bf r}^2
\frac{\partial}{\partial {\bf r}}-2{\bf X}({\bf X}{\bf P})-\nonumber\\
&&\frac{2m_1m_2}{(m_1+m_2)^2}{\bf r}({\bf r}{\bf P})+
2\imath {\bf X}({\bf r}\frac{\partial}{\partial {\bf r}})+
2\imath {\bf r}({\bf X}\frac{\partial}{\partial {\bf r}})+\nonumber\\
&&\frac{2\imath (m_2-m_1)}{m_1+m_2}
{\bf r}({\bf r}\frac{\partial}{\partial {\bf r}}) +6\imath {\bf X}+
\frac{3\imath (m_2-m_1)}{m_1+m_2}{\bf r}]+\nonumber\\
&&\frac{1}{R}[({\bf s}_1+{\bf s}_2)\times {\bf X}+
\frac{m_2{\bf s}_1-m_1{\bf s}_2}{m_1+m_2}\times {\bf r}]
\label{19}
\end{eqnarray}
where ${\bf S}={\bf l}({\bf r})+{\bf s}_1+{\bf s}_2$ and
${\bf P}=-\imath \partial/\partial {\bf X}$.

 A direct calculation shows that, as a consequence of Eqs. (\ref{11}),
(\ref{16}) and (\ref{19}),
\begin{eqnarray}
&& M_{dS}^2=[\mu_1+\mu_2-\imath({\bf r}\frac{\partial}{\partial
{\bf r}}+\frac{3}{2})]^2+\frac{\mu_1\mu_2}{(\mu_1+\mu_2)^2}
[{\bf r}^2{\bf P}^2-\nonumber\\
&& 2({\bf r}{\bf P})^2]+\frac{\imath (\mu_2-m_1)}{\mu_2+\mu_1}
[2({\bf r}{\bf P})({\bf r}\frac{\partial}{\partial
{\bf r}})+3{\bf r}{\bf P}-
{\bf r}^2({\bf P}\frac{\partial}{\partial {\bf r}})]+ \nonumber\\
&&\frac{4(\mu_2{\bf s}_1-\mu_1{\bf s}_2)({\bf r}\times {\bf P})}
{\mu_1+\mu_2}
\label{dS}
\end{eqnarray}
\begin{eqnarray}
&& M^2=[m_1+m_2-\frac{\imath}{R}({\bf r}\frac{\partial}{\partial
{\bf r}}+\frac{3}{2})]^2+\frac{m_1m_2}{R^2(m_1+m_2)^2}
[{\bf r}^2{\bf P}^2-\nonumber\\
&& 2({\bf r}{\bf P})^2]+\frac{\imath (m_2-m_1)}{R^2(m_2+m_1)}
[2({\bf r}{\bf P})({\bf r}\frac{\partial}{\partial
{\bf r}})+3{\bf r}{\bf P}-
{\bf r}^2({\bf P}\frac{\partial}{\partial {\bf r}})]+ \nonumber\\
&&\frac{4(m_2{\bf s}_1-m_1{\bf s}_2)({\bf r}\times {\bf P})}
{R^2(m_1+m_2)}
\label{20}
\end{eqnarray}
The expressions for both $M_{dS}$ and $M$ have been explicitly
written down in order to stress that if $M_{dS}$ is expressed
in terms of the de Sitter masses then it does not depend on $R$.
Such a dependence arises only when we consider the conventional
mass operator in terms of conventional masses
(see the discussion in the preceding section).

 As follows from Eq. (\ref{20}), the decomposition of the tensor
product of two UIRs belonging to the principal series contains not
only UIRs belonging to this series. Indeed, the spectrum of the
operator $M^2$ is not positive definite. This is clear, for example,
from the fact that for very large values of $|{\bf P}|$ and the
values of ${\bf r}$ collinear with ${\bf P}$, the value of $M^2$
becomes negative. However if $R$ is very large then such values of
$|{\bf P}|$ are practically impossible. It is obvious from Eq.
(\ref{20}) that for realistic values of $|{\bf P}|$ the contribution
to $M^2$ of the UIRs not belonging to the principal series is a
small correction of order $1/R^2$.

 If only the contribution of UIRs belonging to the principal
series is taken into account, then the problem of determining
$M^2$ can be considered as follows. Since the tensor product
of two UIRs can be decomposed into the direct integral of UIRs
\cite{Dix} and any UIR of the principal series is unitarily
equivalent to the representation (\ref{10}) with some operators
${\bf s}$ and values of $\mu$, we can conclude that any
representation of the SO(1,4) group containing only the UIRs of
the principal series is unitarily equivalent to the representation
defined by the generators
\begin{eqnarray}
&&{\bf M}={\bf l}({\bf X})+{\bf S}, \quad L_{+-}=-2(M_{dS}+{\bf X}{\bf P})+
3\imath,\nonumber\\
&& {\bf Q}_+=-2R{\bf P}, \quad {\bf Q}_-=\frac{1}{2R}[-2M_{dS} {\bf X}+
{\bf X}^2{\bf P}- \nonumber\\
&& 2{\bf X}({\bf X}{\bf P})+3\imath {\bf X} +2({\bf S}\times {\bf X})]
\label{21}
\end{eqnarray}
where the generators ${\bf S}$ and $M_{dS}$ act only through the
internal variables of the system under consideration. By analogy
with the case of UIRs considered in the preceding section, it is
clear that $M_{dS}$ can be determined by considering the action
of $L_{+-}$ on the states with ${\bf P}=0$: the action of $L_{+-}$
on such states is equal to $2(-M_{dS}+3\imath /2)$ (recall that
the sign of $M_{dS}$ does not play a role). Therefore, as follows
from Eq. (\ref{19}), the mass operator in the given case is
unitarily equivalent to the operator
\begin{equation}
M=m_1+m_2-\frac{\imath}{R}({\bf r}\frac{\partial}{\partial {\bf r}}+
\frac{3}{2})
\label{22}
\end{equation}
In particular, the positive part of the operator (\ref{20}) is
unitarily equivalent to the square of the operator given by
Eq. (\ref{22}). This is in agreement with the "common wisdom"
according to which the spectrum of the mass operator is defined
by its reduction on the (generalized) subspace of states with
${\bf P}=0$.

 The comparison of Eqs. (\ref{17}) and (\ref{22}) is an additional
argument in treating $S^3(R_1)$ as the coordinate space, at least
at very small velocities (note that
${\bf r}=\imath \partial/\partial {\bf k}$,
${\bf k}=-\imath \partial/\partial {\bf r}$). We will use momentum
and coordinate representations depending on convenience. In the
first case the operator (\ref{17}) acts in the space of functions
$f({\bf k})$ such that
\begin{equation}
\int\nolimits |f({\bf k})|^2 d^3{\bf k} < \infty
\label{23}
\end{equation}
and in the second case --- in the space of functions
$\varphi ({\bf r})$ such that
\begin{equation}
\int\nolimits |\varphi ({\bf r})|^2 d^3{\bf r} < \infty
\label{24}
\end{equation}
(here and henceforth we will consider only the spinless case
for simplicity). The functions
$f({\bf k})$ and $\varphi ({\bf r})$ are the Fourier transforms of
each other.

 In spherical coordinates ${\bf r} \partial/\partial {\bf r}
=r \partial/\partial r$ where
$r=|{\bf r}|$. Therefore in these coordinates the operator (\ref{22})
does not act through the angular variables. We can consider the
action of this operator in the space of functions $\varphi (r)$
such that
\begin{equation}
\int\nolimits |\varphi (r)|^2 r^2dr < \infty
\label{25}
\end{equation}
The eigenvalue problem
\begin{equation}
(m_1+m_2)\varphi_{\lambda}(r)-\frac{\imath}{R}
[r\frac{d\varphi_{\lambda} (r)}{dr}
+\frac{3}{2}\varphi_{\lambda} (r)]=\lambda \varphi_{\lambda} (r)
\label{26}
\end{equation}
has the solution
\begin{equation}
\varphi_{\lambda} (r)=\frac{1}{r}(\frac{R}{2\pi r})^{1/2}exp[\imath R
(\lambda-m_1-m_2)ln\,r]
\label{27}
\end{equation}
where we assume that $r$ is given in some dimensional units. Then
\begin{eqnarray}
&& \int_{0}^{\infty} \varphi_{\lambda} (r)^*\varphi_{\lambda'} (r)
r^2dr =\delta (\lambda-\lambda'),\nonumber\\
&& \int_{-\infty}^{\infty} \varphi_{\lambda} (r)^*\varphi_{\lambda} (r')
d\lambda =\frac{1}{r^2}\delta (r-r')
\label{28}
\end{eqnarray}
where $^*$ means the complex conjugation.
Therefore the operator (\ref{22}) contains only the continuous
spectrum occupying the interval $(-\infty,\infty)$. The same is
obviously valid for the operator (\ref{17}). As a consequence, we have

 {\it Statement 1}: {\it The operators} (\ref{17}) {\it and} (\ref{22})
{\it are unitarily equivalent to the operator}
\begin{equation}
M_0=\frac{\imath}{R}({\bf k}\frac{\partial}{\partial {\bf k}}+
\frac{3}{2})=-\frac{\imath}{R}({\bf r}\frac{\partial}{\partial {\bf r}}+
\frac{3}{2})
\label{29}
\end{equation}

 When $R\rightarrow \infty$ we must have the compatibility of the
above results with the standard results of the Poincare invariant
theory, according to which the mass operator is given by
$M^{P}=(m_1^2+k^2)^{1/2}+(m_2^2+k^2)^{1/2}$ where $k=|{\bf k}|$. We use
$g(k^2)$ to denote the function
\begin{eqnarray}
g(k^2)&=&\frac{1}{2}\int_{0}^{k^2}\{[(m_1^2+
k^{'2})^{1/2}+m_1]^{-1}+\nonumber\\
&&[(m_2^2+k^{'2})^{1/2}+m_2]^{-1}\}dk^{'2}
\label{30}
\end{eqnarray}
Then it is obvious that
\begin{eqnarray}
&& (m_1^2+k^2)^{1/2}+(m_2^2+k^2)^{1/2}+\frac{\imath}{R}({\bf k}
\frac{\partial}{\partial {\bf k}}+\frac{3}{2})=\nonumber\\
&&exp[\imath Rg(k^2)][m_1+m_2+\frac{\imath}{R}({\bf k}
\frac{\partial}{\partial {\bf k}}+
\frac{3}{2})]\nonumber\\
&&exp[-\imath Rg(k^2)]
\label{31}
\end{eqnarray}
We can formulate this result as

{\it Statement 2}: {\it The operator}
\begin{equation}
{\tilde M}=(m_1^2+k^2)^{1/2}+(m_2^2+k^2)^{1/2}+\frac{\imath}{R}({\bf k}
\frac{\partial}{\partial {\bf k}}+\frac{3}{2})
\label{32}
\end{equation}
{\it is unitarily equivalent to the operator given by Eq.} (\ref{17}).

If one considers the action of the operator ${\tilde M}$ on functions
$f({\bf k})$ satisfying the conditions
\begin{equation}
|\partial f({\bf k})/\partial {\bf k}|\ll R|f({\bf k})|
\label{33}
\end{equation}
then the action of ${\tilde M}$ is practically indistinguishable
from the action of $M^P$.

 In the approximation when only the representations containing the
UIRs of the principal series are considered, the above results can
be easily generalized to the case of any number of particles.
Indeed, suppose that the system under consideration consists of
two subsystems, $\alpha$ and $\beta$. Let $M_{\alpha}$ and
$M_{\beta}$ be the corresponding mass operators and
${\bf k}_{\alpha\beta}$ be the relative momentum. Then, by analogy
with Eq. (\ref{17}),
\begin{equation}
M_{\alpha\beta}=M_{\alpha}+M_{\beta}+\frac{\imath}{R}
({\bf k}_{\alpha\beta}\frac{\partial}{\partial {\bf k}_{\alpha\beta}}+
\frac{3}{2})
\label{34}
\end{equation}

 By analogy with the above consideration it is easy to show that
this operator is unitarily equivalent to
\begin{equation}
{\tilde M}_{\alpha\beta}=(M_{\alpha}^2+k_{\alpha\beta}^2)^{1/2}+
(M_{\beta}^2+k_{\alpha\beta}^2)^{1/2}+\frac{\imath}{R}
({\bf k}_{\alpha\beta}\frac{\partial}{\partial {\bf k}_{\alpha\beta}}+
\frac{3}{2})
\label{35}
\end{equation}
where $k_{\alpha\beta}=|{\bf k}_{\alpha\beta}|$. Therefore when
$R$ is large, the Hilbert space of the internal wave functions
contains a subspace of functions with the following property:
the action of ${\tilde M}_{\alpha\beta}$ on these functions is
practically indistinguishable from the action of the standard mass
operator $(M_{\alpha}^2+k_{\alpha\beta}^2)^{1/2}+
(M_{\beta}^2+k_{\alpha\beta}^2)^{1/2}$. It is also clear that
the cosmological repulsion has a place for any pair of particles
in the system under consideration.

 Consider, for example, a system of three particles with the masses
$m_1$, $m_2$ and $m_3$, and introduce the standard Jacobi
variables
\begin{equation}
{\bf k}_{12}=\frac{m_2{\bf p}_1-m_1{\bf p}_2}{m_1+m_2},\quad
{\bf K}_{12}=\frac{m_3({\bf p}_1+{\bf p}_2)-(m_1+m_2){\bf p}_3}
{m_1+m_2+m_3}
\label{36}
\end{equation}
Then it follows from Eqs. (\ref{17}) and (\ref{34}) that
\begin{equation}
M_{123}=m_1+m_2+m_3+\frac{\imath}{R} ({\bf k}_{12}\frac{\partial}
{\partial {\bf k}_{12}}+\frac{3}{2})+
\frac{\imath}{R} ({\bf K}_{12}\frac{\partial}
{\partial {\bf K}_{12}}+\frac{3}{2})
\label{37}
\end{equation}
This operator is unitarily equivalent to
\begin{eqnarray}
&&{\tilde M}_{123}=(M_{12P}^2+K_{12}^2)^{1/2}+(m_3^2+
K_{12}^2)^{1/2}+ \nonumber\\
&&\frac{\imath}{R} ({\bf k}_{12}\frac{\partial}
{\partial {\bf k}_{12}}+\frac{3}{2})+
\frac{\imath}{R} ({\bf K}_{12}\frac{\partial}
{\partial {\bf K}_{12}}+\frac{3}{2})
\label{38}
\end{eqnarray}
where $M_{12}^P$ is the mass operator of the system (12) in the
Poincare invariant theory. There exists a subspace
of functions $f({\bf k}_{12},{\bf K}_{12})$ with the following
property: the action of ${\tilde M}_{123}$ on these functions is
practically indistinguishable from the action of the standard mass
operator $(M_{12P}^2+K_{12}^2)^{1/2}+
(m_3^2+K_{12}^2)^{1/2}$. The functions
$f({\bf k}_{12},{\bf K}_{12})$ should satisfy the property
\begin{equation}
|\frac{\partial f({\bf k}_{12},{\bf K}_{12})}
{\partial {\bf k}_{12}}|,|\frac{\partial f({\bf k}_{12},
{\bf K}_{12})}{\partial {\bf K}_{12}}|\ll R|f({\bf k}_{12},
{\bf K}_{12})|
\label{39}
\end{equation}

\begin{sloppypar}
\section{Unitary equivalence of free and interacting representations
in the SO(1,4)-invariant theory}
\label{S4}
\end{sloppypar}

  If the particles in the system under consideration interact with
each other then the representation generators describing this
system are interaction dependent but they must satisfy the
commutation relations (\ref{2}). By analogy with the procedure
proposed by Bakamdjian and Thomas in Poincare-invariant theories
\cite{BT}, we can introduce the interaction by replacing the free
mass operator $M_{dS}$ in Eq. (\ref{21}) by an interacting mass
operator ${\hat M}_{dS}$. Then the relations (\ref{21}) will be
obviously satisfied if ${\hat M}_{dS}$ acts only through the
internal variables and commutes with ${\bf S}$.

 In the general case the spin and momentum operators in Eq. (\ref{21})
can be interaction dependent too but, by analogy with
Poincare-invariant theories (see e.g. Ref. \cite{lev1}) it is natural to
assume that they have the same spectrum as the corresponding free
operators. Therefore one can eliminate the interaction dependence
of the spin and momentum operators by using a proper unitary
transformation. In Poincare-invariant theories the corresponding
unitary operators are known as Sokolov's packing operators
(see e.g. Refs. \cite{sok,CP,Mutze,lev1,KeiPol,lev2}).

 In the present paper we will consider the representation generators
in the Bakamdjian-Thomas (BT) form, but the above discussion
gives grounds to think that any representation describing the
interacting system in the SO(1,4)-invariant theory and taking into
account only the contribution of the UIRs of the principal series
is unitarily equivalent to the representation in the BT form.

 Consider first in detail the case of two free particles in the
nonrelativistic approximation. We can write ${\tilde M}-m_1-m_2$ in
the form
\begin{equation}
M_{nr}=\frac{k^2}{2m_{12}}+\frac{\imath}{R} ({\bf k}\frac{\partial}
{\partial {\bf k}}+\frac{3}{2})
\label{40}
\end{equation}
where $m_{12}=m_1m_2/(m_1+m_2)$ is the reduced mass of particles
1 and 2. In coordinate representation this operator has the form
\begin{equation}
M_{nr}=-\frac{\Delta}{2m_{12}}-\frac{\imath}{R} ({\bf r}\frac{\partial}
{\partial {\bf r}}+\frac{3}{2})
\label{41}
\end{equation}
where $\Delta =(\partial /\partial {\bf r})^2$.

  It is obvious that
\begin{equation}
M_{nr}=exp(\frac{\imath Rk^2}{4m_{12}})M_0
exp(-\frac{\imath Rk^2}{4m_{12}})
\label{42}
\end{equation}
where $M_0$ is given by Eq. (\ref{29}). Therefore we have

{\it Statement 3}: {\it The operators} $M_0$ {\it and} $M_{nr}$
{\it are unitarily equivalent}. In particular, $M_{nr}$ contains
only the continuos spectrum occupying the interval $(-\infty,\infty)$.

The Hilbert space of functions satisfying the conditions (\ref{23}) or
(\ref{24}) can be decomposed into the subspaces $H_{lm}$ such that the
elements of $H_{lm}$ have the form
\begin{equation}
f({\bf k})=Y_{lm}({\bf k}/k)f(k),\quad
\varphi ({\bf r})=Y_{lm}({\bf r}/r)\varphi (r)
\label{43}
\end{equation}
$Y_{lm}$ is the spherical function, $\varphi (r)$ satisfies the
condition (\ref{25}) and $f(k)$ satisfies the analogous condition
in momentum representation.

In this representation the eigenvalue
problem for the operator $M_{nr}$ in $H_{lm}$ does not depend on $l$
and $m$:
\begin{equation}
\frac{k^2}{2m_{12}}f_{\lambda}(k)+\frac{\imath}{R}
[k\frac{df_{\lambda}(k)}{k}+\frac{3}{2}f_{\lambda}(k)]=\lambda
f_{\lambda}(k)
\label{44}
\end{equation}
The solution of this equation is
\begin{equation}
f_{\lambda}(k)=\frac{1}{k}(\frac{R}{2\pi k})^{1/2}exp[\imath R
(\frac{k^2}{4m_{12}}-\lambda ln\,k)]
\label{45}
\end{equation}

 In coordinate representation the eigenvalue
problem for the operator $M_{nr}$ in $H_{lm}$ reads
\begin{eqnarray}
&&-\frac{1}{2m_{12}r^2}\frac{d}{dr}[r^2\frac{d\varphi_{\lambda l} (r)}{dr}]+
\frac{l(l+1)}{2m_{12}r^2}\varphi_{\lambda l} (r)-\nonumber\\
&&-\frac{\imath}{R}
[r\frac{d\varphi_{\lambda l} (r)}{dr}
+\frac{3}{2}\varphi_{\lambda l} (r)]=\lambda \varphi_{\lambda l} (r)
\label{46}
\end{eqnarray}

 The relation between the functions $f_{\lambda}(k)$ and
$\varphi_{\lambda l} (r)$ is given by the radial Fourier transform:
\begin{equation}
\varphi_{\lambda l} (r)=\frac{R^{1/2}}{\pi} (-\imath )^l
\int_{0}^{\infty} j_l(kr) k^{1/2} exp[\imath R
(\frac{k^2}{4m_{12}}-\lambda ln\,k)] dk
\label{47}
\end{equation}
where
\begin{equation}
j_l(kr)=(\frac{\pi}{2kr})^{1/2}J_{l+1/2}(kr)
\label{48}
\end{equation}
is the spherical Bessel function.

 The integral in Eq. (\ref{47}) can be calculated analytically. We use
$\gamma^2$ to denote $-\imath R/4m_{12}$. Then \cite{BE}
\begin{eqnarray}
&& \varphi_{\lambda l} (r)=(\frac{R}{2\pi r})^{1/2}
\frac{(-\imath )^l}{2\Gamma (l+3/2)}
\gamma^{\imath \lambda R -1}(\frac{r}{2\gamma})^{l+1/2}\nonumber\\
&&\Gamma (\frac{l}{2}+\frac{3}{4} -\frac{\imath \lambda R}{2})
_1F_1(\frac{l}{2}+\frac{3}{4} -\frac{\imath \lambda R}{2};
l+\frac{3}{2};\frac{r^2}{4\gamma^2})
\label{49}
\end{eqnarray}
where $\Gamma$ is the gamma function and $_1F_1$ is the hypergeometric
function (in Ref. \cite{BE} a general case is considered and one has
to require $Re(\gamma^2)>0$ but in our case it is also possible to use
the result of Ref. \cite{BE} if $Re(\gamma^2)=0$).

 As follows from Eq. (\ref{49}), when $r\rightarrow 0$, the
function $\varphi_{\lambda l} (r)$ is proportional to $r^l$.
This fact was clear from the analogy with the conventional
quantum mechanics. Indeed, by analogy with the standard
investigation of the radial Schroedinger equations, one can expect that
at $r\rightarrow 0$ the first two terms in Eq. (\ref{46}) are
dominant. It is well-known that the only regular solution at
such conditions is proportional to $r^l$.

 Let us now consider the asymptotic behavior of
$\varphi_{\lambda l} (r)$ when $r\rightarrow \infty$. It is
convenient to use not Eq. (\ref{49}) but the original integral
(\ref{47}). Introducing the new integration variable $t=kr$ and
using Eq. (\ref{48}) one arrives at the following asymptotic
result:
\begin{eqnarray}
\varphi_{\lambda l} (r)&=&\frac{(-\imath)^l}{r}(\frac{R}{2\pi r})^{1/2}
exp(\imath \lambda R)\nonumber\\
&&\int_{0}^{\infty} J_{l+1/2}(t)exp(-\imath R\lambda ln\,t)dt
\label{50}
\end{eqnarray}
Since \cite{BE}
\begin{eqnarray}
&&\int_{0}^{\infty} J_{l+1/2}(t)exp(-\imath R\lambda ln\,t)dt=\nonumber\\
&&2^{-\imath R\lambda}
\Gamma (\frac{l}{2}+\frac{3}{4} -\frac{\imath \lambda R}{2})/
\Gamma (\frac{l}{2}+\frac{3}{4} +\frac{\imath \lambda R}{2})
\label{51}
\end{eqnarray}
the comparison of Eqs. (\ref{46}), (\ref{50}) and (\ref{51})
with Eqs. (\ref{26}) and (\ref{27}) shows that at
$r\rightarrow \infty$ one can neglect the first two terms in
Eq. (\ref{46}).

We can normalize the functions $\varphi_{\lambda l} (r)$ as
\begin{equation}
\int_{0}^{\infty}\varphi_{\lambda l} (r)^*\varphi_{\lambda l} (r)
r^2dr=\delta (\lambda -\lambda')
\label{52}
\end{equation}
and any function $\varphi ({\bf r})$ from the internal Hilbert
space can be written as
\begin{equation}
\varphi ({\bf r})=\sum_{lm}\int_{-\infty}^{\infty} c_{lm}(\lambda)
Y_{lm}({\bf r}/r)\varphi_{\lambda l} (r)d\lambda
\label{53}
\end{equation}

Now we proceed to the case of interacting particles and consider the
operator
\begin{equation}
{\hat M}_{nr}=-\frac{\Delta}{2m_{12}}+V(r)-
\frac{\imath}{R} ({\bf r}\frac{\partial}{\partial {\bf r}}+\frac{3}{2})
\label{54}
\end{equation}
The eigenvalue problem for this operator in $H_{lm}$ has the form
\begin{eqnarray}
&&-\frac{1}{2m_{12}r^2}\frac{d}{dr}[r^2\frac{d\psi_{\lambda l} (r)}{dr}]+
\frac{l(l+1)}{2m_{12}r^2}\psi_{\lambda l} (r)+V(r)
\psi_{\lambda l} (r)-\nonumber\\
&&\frac{\imath}{R} [r\frac{d\psi_{\lambda l} (r)}{dr}
+\frac{3}{2}\psi_{\lambda l} (r)]=\lambda \psi_{\lambda l} (r)
\label{55}
\end{eqnarray}

 Suppose that $V(r)r^2\rightarrow 0$ when $r\rightarrow 0$. Then
the third term in Eq. (55) is negligible in comparison with the
first two ones when $r\rightarrow 0$ (see e.g. Ref. \cite{LL}).
Therefore the asymptotic behavior of the function
$\psi_{\lambda l} (r)$ at $r\rightarrow 0$ is the same as that
of the function $\varphi_{\lambda l} (r)$, i.e.
$\psi_{\lambda l} (r)$ is proportional to $r^l$. On the other
hand, if $V(r)\rightarrow 0$ when $r\rightarrow \infty$ then
the third term in Eq. (55) is negligible in comparison with the
fourth one when $r\rightarrow \infty$.
Therefore the asymptotic behavior of the function $\psi_{\lambda l} (r)$
at $r\rightarrow \infty$ also coincides with that of the function
$\varphi_{\lambda l} (r)$, i.e. the function $\psi_{\lambda l} (r)$
at $r\rightarrow \infty$ is proportional to the expression given
by Eq. (\ref{50}).

  Since the functions $\psi_{\lambda l} (r)$ and
$\varphi_{\lambda l} (r)$ have the same asymptotic behavior at
$r\rightarrow 0$ and $r\rightarrow \infty$, we conclude that
the operators $M_{nr}$ and ${\hat M}_{nr}$ have the same spectrum.
Since the normalization of the eigenfunctions belonging to the
continuous spectrum is fully determined by the asymptotic behavior
of these functions at $r\rightarrow \infty$ \cite{LL}, we can
normalize the functions $\psi_{\lambda l} (r)$ in the same way
as the functions $\varphi_{\lambda l} (r)$:
\begin{equation}
\int_{0}^{\infty}\psi_{\lambda l} (r)^*\psi_{\lambda l} (r)
r^2dr=\delta (\lambda -\lambda')
\label{56}
\end{equation}
Then we can define the operator $U$ as follows. If the
function $\varphi ({\bf r})$ is given by Eq. (\ref{53}) then
\begin{equation}
U\varphi ({\bf r})=\sum_{lm}\int_{-\infty}^{\infty} c_{lm}(\lambda)
Y_{lm}({\bf r}/r)\psi_{\lambda l} (r)d\lambda
\label{57}
\end{equation}
The operator $U$ commutes with ${\bf S}$ by construction.
As follows from Eqs. (\ref{52}) and (\ref{56}), this operator
is unitary, and ${\hat M}_{nr}=UM_{nr}U^{-1}$.
Therefore we have

{\it Statement 4: The operators} ${\hat M}_{nr}$ {\it and} $M_{nr}$
{\it are unitarily equivalent}.

  In the nonrelativistic approximation we have to consider functions
$f({\bf k})$ for which the important values of ${\bf k}$ are much
smaller than the masses of the particles in question. Suppose also
that $R$ is very large and the functions satisfy the condition
(\ref{33}). In coordinate representation the actions of $M_{nr}$
and ${\hat M}_{nr}$ on such functions are practically
indistinguishable from the actions of the operators
\begin{equation}
M_{nr}^{G}=-\frac{\Delta}{2m_{12}}\quad \mbox{and} \quad
{\hat M}_{nr}^{G}=-\frac{\Delta}{2m_{12}}+V(r),
\label{58}
\end{equation}
respectively ("G" means "Galilei"). However the operators
$M_{nr}^{G}$ and ${\hat M}_{nr}^{G}$ are not necessarily
unitarily equivalent. For example, if $V(r)=-const/r$ and
$const>0$ then the
operator $M_{nr}^{G}$ has only the continuous spectrum in the
interval $[0,\infty)$ while ${\hat M}_{nr}^{G}$ has also the
discrete spectrum at some negative values of $\lambda$. Since the
operators $M_{nr}^{G}$ and ${\hat M}_{nr}^{G}$ in this case have
different spectra, they cannot be unitarily equivalent.

 The result formulated in {\it Statement 4} could be expected from
physical considerations. Indeed, if $V(r)$ is not too singular when
$r\rightarrow 0$, the phenomenon known as the "fall onto the
center" (see e.g. Ref. \cite{LL}) does not occur and the spectra
of $M_{nr}$ and ${\hat M}_{nr}$ are defined by asymptotic of the
eigenfunctions of these operators at $r\rightarrow \infty$. Even
if $R$ is very large, there exist such values of $r$ that the
cosmological repulsion becomes dominant in comparison with the
kinetic and potential energies. Since this repulsion is present in
both $M_{nr}$ and ${\hat M}_{nr}$, these operators have the same
spectrum and therefore are unitarily equivalent.

 Consider now a system of $N$ particles with arbitrary velocities
and suppose that the interactions between the particles can be
described only in terms of the degrees of freedom characterizing
these particles. The gravitational and electromagnetic interactions
are not very singular in the sense that they
fall off at infinity and do not lead to the fall onto the center
\cite{LL}. In the case of strong interactions the problem exists
how to describe the interaction of colored objects at large
distances. Such an interaction is often modelled by attractive
potentials which at infinity are proportional to $r$. In this case
the force of attraction does not depend on $r$ and therefore can
be neglected in comparison with the cosmological repulsion.
Therefore it is natural to say that at infinity all realistic
interactions are negligible in comparison with the cosmological
repulsion (or by definition, the necessary condition for any
interaction to be realistic is to be small in comparison with
the cosmological repulsion when $r\rightarrow\infty$).
For these reasons, in the case of realistic interactions, asymptotic
of the eigenfunctions of the interacting mass operator is again
defined by the cosmological repulsion and is the same as asymptotic
of the eigenfunctions of the free mass operator discussed in the
preceding section. Therefore we can formulate the following

{\it Statement 5: In the SO(1,4)-invariant theory the interacting
mass operator of the system of} $N$ {\it particles described by the
UIRs of the principal series is unitarily equivalent to the free
mass operator}.

 If the unitary operator realizing the equivalence of two mass
operators commutes with ${\bf S}$ and the corresponding representations
can be realized in the BT form then they are unitarily equivalent.

 In QED, electroweak theory and QCD the Hilbert space for the
system under consideration is the Fock space describing a system of
infinite number of particles. As noted in Sect. \ref{S1}, there
exist serious mathematical problems in constructing the representation
operators of the Poincare group in these theories. The problem
becomes much more complicated if the symmetry group is the group of
motions of a curved space-time (see e.g. Ref. \cite{BirDav}).
The above considerations give grounds to think that in all
SO(1,4)-invariant theories of realistic interactions the spectrum of
the representation generators is fully defined by the cosmological
repulsion which is dominant at large distances. Therefore we can
formulate the following

{\it Conjecture: In any SO(1,4)-invariant theory of realistic
interactions the interacting and free representation generators are
unitarily equivalent.}

\section{Discussion}
\label{S5}

 A standard problem of the perturbation theory for linear operators
(see e.g. Ref. \cite{Kato}) is as follows. Let $A$ and ${\hat A}$ be
selfadjoint operators in the Hilbert space. Suppose that they have
the same (absolutely) continuous spectrum (this is treated as the
property of ${\hat A}$ to be in some sense a small perturbation of $A$).
Suppose also for simplicity that $A$ does not contain other points
of the spectrum. Consider the wave operators
\begin{eqnarray}
W_{\pm}(A,{\hat A})=s-lim_{t \rightarrow \pm\infty}\,\,
exp(\imath {\hat A}t)exp(-\imath At)
\label{60}
\end{eqnarray}
where $s-lim$ means the strong limit. If ${\hat A}$ has the same spectrum
as $A$ then there exist conditions when the operators $W_{\pm}$ are
unitary and
\begin{equation}
{\hat A}=W_{\pm}(A,{\hat A})AW_{\pm}(A,{\hat A})^{-1}
\label{61}
\end{equation}
i.e. $A$ and ${\hat A}$ are unitarily equivalent. If ${\hat A}$ also
contains the
discrete spectrum, the operators $A$ and ${\hat A}$ cannot be unitarily
equivalent but there exist conditions
when the operators $W_{\pm}$ are isometric and the $S$-operator
$S=W_+^*W_-$ is unitary.

 As shown in the preceding section, in the SO(1,4)-invariant theory
the interacting mass operator ${\hat M}$ of a many-particle system
has the same spectrum as the free mass operator $M$ and these
operators are unitarily equivalent. The absence of bound states is
a consequence of the fact that at large distances the cosmological
repulsion is dominant. The choice of the unitary operator realizing
the equivalence of ${\hat M}$ and $M$ is obviously not unique. By
analogy with the standard results of the perturbation theory for
linear operators one could expect that a possible choice is
$W_{\pm}(M,{\hat M})$.

 If $R$ is very large and one considers only the subspace of
$H^P$ of functions satisfying the conditions analogous to
(\ref{33}), (\ref{39}) etc. then the actions of ${\hat M}$ and $M$
on these functions are practically indistinguishable from the
actions of the corresponding operators in the Poincare-invariant
theory (obtained from $M$ and ${\hat M}$ by neglecting the
cosmological repulsion). Therefore in the SO(1,4)-invariant
theory there exist quasi-bound states: their lifetime is very
large and goes to infinity when $R\rightarrow\infty$. It is clear
that there exist conditions when the quasi-bound states are
practically indistinguishable from bound ones. The finite lifetime
is related to the fact that theoretically there exists a nonzero
probability for quasi-bound particles to pass through the barrier
separating the usual and cosmological distances. However in practice
this probability can be negligible.

 It is important to note that the subspace $H^P$, where the results
of the Poincare-invariant theory are valid, is only a small part of
the full Hilbert space $H$. In particular, if $f\in H^P$ then
$exp(\imath {\hat M}t)$ and $exp(-\imath Mt)$ cannot belong to
$H^P$ if $t$ is of order $R/c$. Therefore it is natural to think that
the standard scattering problem in $H^P$ is meaningful only if
$ct\ll R$.

 The above remarks may also be considered as an indication that
difficulties of the present quantum theory (e.g. divergencies) are
related to the fact that loop contributions to the $S$-operator
involve not only a set of states belonging to $H^P$ but a much
wider set of states for which it is not possible to neglect
the effects of de Sitter invariance. In other words, these effects
can play a role of regularizers for usual divergent expressions
in the standard approach. For the case of the SO(2,3) symmetry
such a possibility has been considered in Ref. \cite{Kad}.

 If {\it Conjecture} formulated in the preceding section is valid
then it is natural to think that the very notion of interaction is
not fundamental. Indeed, in this case {\it there is no need to
introduce interaction terms into the operators: we can always
work with the free operators and physics is defined by the subset
of states important in the processes under consideration}.

 The subsets corresponding to different interactions are connected
with each other by unitary transformations which necessarily
depend on $R$. Indeed, if one reduces the free and interacting
operators onto $H^P$ and neglects the cosmological repulsion, then,
as noted above, the operators obtained in such a way are not
unitarily equivalent in the general case.

 In particular, one might think that for the situation
corresponding to the pair of the operators
$M_{nr}$ and ${\hat M}_{nr}$ (see the preceding section) the
really fundamental problem is not the choice of the potential
$V(r)$ which should be added to $M_{nr}$ but the choice of the
unitary operator $U$ realizing the unitary equivalence of
$M_{nr}$ and ${\hat M}_{nr}$. In the framework of such an approach
one might think that the really fundamental quantities are those
defining the operator $U$. In this case the gravitational
constant, electric charges etc. are functions of more fundamental
quantities and $R$, in agreement with the famous Dirac hypothesis
\cite{Dir} about the dependence of physical constants on
cosmological parameters.

 Moreover, in view of the above discussion it is natural to think
that all the existing interactions are fully defined by the present
state vector of the Universe. This can be treated as the quantum
analog of Mach's principle according to which the local physical
laws are defined by the distribution of masses in the Universe
(the discussion of Mach's principle and its relation to general
relativity and Dirac's cosmology can be found in wide literature
--- see e.g. Ref. \cite{Mach} and references therein).

 In summary, SO(1,4)-invariant theories have
rather unusual properties, in particular the mass operator has
only continuous spectrum in the interval $(-\infty,\infty)$,
bound states do not exist and the representations describing free
and interacting systems are unitarily equivalent. At the same time
SO(1,4) invariance does not contradict the existing experimental
data and therefore there exists the possibility that the SO(1,4)
group is the symmetry group of the nature. For these reasons
the investigation of SO(1,4)-invariant theories is of indubitable
interest.

\begin{center} {\it Acknowledgment} \end{center}

 The author is grateful to E. Pace, G. Salme, A.N. Sissakyan and
E. Tagirov for useful discussions. This work was supported in part
by the grant 96-02-16126a from the Russian Foundation for Basic
Research.


\begin{thebibliography}{99}
\bibitem{Wheeler} J.A. Wheeler, {\it Neutrinos, Gravitation and
Geometry} (Scuola Internazionale di Fisica "Enrico Fermi",
Bologna, 1960).
\bibitem{time} W. Pauli, {\it Handbuch der Physik}, vol. V/1
(Berlin, 1958); F.T. Smith, Phys. Rev. {\bf 118}, 349 (1960);
Y. Aharonov and D. Bohm, Phys. Rev. {\bf 122}, 1649 (1961),
{\bf 134}, 1417 (1964); V.A. Fock, ZhETF {\bf 42}, 1135 (1962);
B.A. Lippman, Phys. Rev. {\bf 151}, 1023 (1966).
\bibitem{Heg} H. Saleker and E.P. Wigner, Phys. Rev. {\bf 109},
571 (1958); G.H. Hegerfeldt and S.N. Ruijsenaars, Phys. Rev.
{\bf D22}, 377 (1980).
\bibitem{Wein} S. Weinberg, {\it The Quantum Theory of Fields}
(Cambridge, Cambridge University Press, 1995).
\bibitem{BLOT} R.F. Streater and A.S. Wightman, {\it PCT, Spin,
Statistics and All That} (W.A. Benjamin Inc. New York-Amsterdam, 1964);
R. Jost, {\it The General Theory of Quantized Fields}, Ed. M. Kac
(American Mathematical Society, Providence, 1965);
N.N. Bogolubov, A.A. Logunov, A.I. Oksak, and I.T. Todorov,
{\it General Principles of Quantum Field Theory} (Nauka, Moscow, 1987).
\bibitem{Zim} W. Zimmerman, Annals of Physics {\bf 77}, 536 (1973);
O.I. Zavialov, {\it Renormalized Feynman Diagrams} (Nauka,
Moscow 1979).
\bibitem{J} R. Jackiw, in {\it Lectures on Current Algebra and its
Applications}, S. Treiman, R. Jackiw and D. Gross, eds. (Princeton
University Press, Princeton NJ 1972).
\bibitem{JL} K. Johnson and F.E. Low, Progr. Theor. Phys. {\bf 37-38},
74 (1966); D.J. Gross and R. Jackiw, Nucl. Phys. {\bf B14}, 269 (1969);
C.G. Callan, S. Coleman and R. Jackiw, Annals of Physics,
{\bf 59}, 42 (1970); J.C. Collins, A. Duncan and S.D. Joglecar, Phys. Rev.
{\bf D16}, 438 (1977); M.A. Shifman, Phys. Repts. {\bf 209}, 341 (1991);
M. Luke, A. Manohar and A. Savage, Phys. Lett. {\bf B288}, 355 (1992);
Xiandong Ji, Phys. Rev. Lett. {\bf 74}, 1071 (1995).
\bibitem{Haag} R. Haag, Kgl. Danske Videnskab. Selsk. Mat.-Fys. Medd.
{\bf 29}, No. 12 (1955).
\bibitem{GSW} M.B. Green, J.H. Schwarz and E. Witten, {\it
Superstring Theory} (Cambridge University Press, Cambridge, 1988).
\bibitem{Per} A.M. Perelomov, Phys. Rep. {\bf 146}, 135 (1987).
\bibitem{BLP} V.B. Berestetskii, E.M. Lifshits and L.P. Pitaevski,
{\it Relativistic Quantum Theory} (Moscow, Nauka, 1973).
\bibitem{Fron} C. Fronsdal and R.B. Haugen, Phys. Rev. {\bf D12}, 3810
(1975).
\bibitem{desitt} W. De Sitter, Bull. Astron. Inst. Netherlands,
{\bf 5}, 211 (1930), {\bf 6}, 141 (1931);
H.P. Robertson and T.W. Noonan, {\it Relativity and Cosmology}
(Saunders, Philadelphia, 1968); R.C. Tolman, {\it
Relativity, Thermodynamics and Cosmology} (Oxford, Clarendon Press,
1969); C.W. Misner, K.S. Thorne and J.A. Wheeler, {\it Gravitation}
(W.H. Freeman and Company, San Francisco, 1973).
\bibitem{Str} S. Stroem, Ann. Inst. Henri Poincare, {\bf 58} 77 (1970).
\bibitem{Men} M.B. Mensky, {\it Method of Induced Representations.
Space-Time and the Concept of Particles} (Nauka, Moscow, 1976).
\bibitem{Moy} P. Moylan, J. Math. Phys. {\bf 24}, 2706 (1983), {\bf 26},
29 (1985).
\bibitem{lev} F.M. Lev, J. Phys. {\bf A21}, 599 (1988); J. Math. Phys.
{\bf 34}, 490 (1993).
\bibitem{Nov} Yu.V. Novozhylov, {\it Introduction to the Theory of
Elementary Particles} (Nauka, Moscow 1972).
\bibitem{lev1} F.M. Lev, J. Phys. {\bf A17}, 2047 (1984).
\bibitem{LM} F.M. Lev and E.G. Mirmovich, VINITI No. 6099, Dep. (1984).
\bibitem{Dix} G.W. Mackey, Ann.Math. {\bf 55}, 101(1952); {\bf 58},
193(1953); M.A. Naimark, {\it Normalized rings} (Nauka, Moscow, 1968);
J. Dixmier, {\it Les  algebres  d'operateurs dans l'espace
   hilbertien} (Gauthier-Villars, Paris, 1969);
   A.O. Barut and R. Raczka, {\it Theory of group representations  and
   applications} (Warsaw: Polish  Scientific  Publishers, 1977).
\bibitem{BT} L.H. Bakamdjian and L.H. Thomas, Phys. Rev. {\bf 92},
1300 (1953).
\bibitem{sok} S.N. Sokolov, Teor. Mat. Fiz. {\bf 36}, 193 (1978);
Doklady Akademii Nauk SSSR {\bf 233}, 575 (1977).
\bibitem{CP} F. Coester and W.N. Polyzou, Phys.Rev. {\bf D26}, 1348
(1982) 1348.
\bibitem{Mutze} U. Mutze, Habilitationsschrift Univ. Munchen (1982);
Phys.Rev. {\bf D29} (1984) 2255.
\bibitem{KeiPol} B.D. Keister and W.N. Polyzou, Adv. Nucl. Part.
Phys. {\bf 21}, 225 (1995).
\bibitem{lev2} F.M. Lev, Phys. Rev. {\bf D49}, 383 (1994);
Annals of Phys. {\bf 237}, 355 (1995).
\bibitem{BE} H. Bateman and A. Erdelyi, {\it Higher transcendental functions,
Vol. 2} (Mc Graw Hill Book Company Inc., New York, 1953).
\bibitem{LL} L.D. Landau, E.M. Lifshits, {\it Quantum mechanics} (Nauka,
Moscow, 1963).
\bibitem{BirDav} N.D. Birrel and P.G.W. Davies, {\it Quantum Fields in
Curved Space} (Cambridge University Press, Cambridge, 1982).
\bibitem{Kato} T. Kato, {\it Perturbation Theory for Linear Operators}
(Springer-Verlag, Berlin-Heidelberg-New York, 1966).
\bibitem{Kad} A.D. Donskov, V.G. Kadyshevsky, M.D. Mateeev and
R.M. Mir-Kasimov, Bulg. Phys. J. {\bf 1}, 58, 233 (1974); {\bf 2},
3 (1975).
\bibitem{Dir} P.A.M. Dirac, Nature {\bf 139}, 323 (1937); Proc.
Roy. Soc. L., {\bf A333}, 403 (1973).
\bibitem{Mach} R.H. Dicke, Nature {\bf 192}, 440 (1961); R.H. Dicke,
in {\it Gravitation and Relativity}, Hong-Yee Chiu and W.F. Hoffman
Eds.  (W.A. Benjamin INC, New York - Amsterdam, 1964); J.A. Wheeler,
{\it ibid};
M. Reinhardt, Ztschr. Naturforsch. {\bf 28a}, 529 (1973);
R. Haller and F. Stadler, {\it Ernst Mach --- Werk und Wirkung}
(Hoelder-Pichler-Tempsky, Wien, 1988).
\end{thebibliography}
\end{document}